\documentclass[12pt]{iopart}
\usepackage{graphicx}
\usepackage{amssymb}

\usepackage{color}
\definecolor{orange}{RGB}{255,127,0}
\definecolor{grey}{RGB}{125,125,125}
\newcommand{\ds}{\displaystyle}

\graphicspath{{c:/fortran/mapas/}}

\begin{document}

\title{Quantifying sudden changes in dynamical systems using symbolic networks}

\author{Cristina Masoller$^1$, Yanhua Hong$^2$, Sarah Ayad$^3$, Francois Gustave$^3$, Stephane Barland$^3$, Antonio J Pons$^1$, Sergio G\'omez$^4$ and Alex Arenas$^4$}

\address{$^1$ Departament de F\'{\i}sica i Enginyeria Nuclear, Universitat Polit\`ecnica de Catalunya, 08222 Terrassa, Spain}

\address{$^2$ School of Electronic Engineering, Bangor University, Bangor, Gwynedd LL57 1UT, Wales, UK.}

\address{$^3$ Universit\'e de Nice Sophia Antipolis, Institut Non-Lin\'eaire de Nice, UMR 6618, 06560 Valbonne, France}

\address{$^4$ Departament d'Enginyeria Inform\`atica i Matem\`atiques, Universitat Rovira i Virgili, 43007 Tarragona, Spain}

\eads{\mailto{cristina.masoller@upc.edu}, \mailto{y.hong@bangor.ac.uk}, \mailto{stephane.barland@inln.cnrs.fr}, \mailto{a.pons@upc.edu}, \mailto{sergio.gomez@urv.cat}, \mailto{alexandre.arenas@urv.cat}}

\begin{abstract}
{\textcolor{black}{We characterise the evolution of a dynamical system by combining two well-known complex systems' tools, namely, symbolic ordinal analysis and networks}}. From the ordinal representation of a time-series we construct a network in which every node weights represents the probability of an ordinal patterns (OPs) to appear in the symbolic sequence and each edges weight represents the probability of transitions between two consecutive OPs. Several network-based diagnostics are then proposed to characterize the dynamics of different systems: logistic, tent and circle maps. We show that these diagnostics are able to capture changes produced in the dynamics as a control parameter is varied. We also apply our new measures to empirical data from semiconductor lasers and show that they are able to anticipate the polarization switchings, thus providing early warning signals of abrupt transitions.
 %Time series analysis; Complex systems; Entropy and other measures of information; Networks and genealogical trees
\end{abstract}

\maketitle

\section{Introduction}

Symbolic time-series analysis is a powerful technique able to extract hidden features such as the presence of frequent recurrent patterns \cite{prl_1989,ijbc_1995,rsi_2003,hutt_prl_2013,depolsi_epjd_2013}, or the existence of missing/forbidden patterns \cite{forbidden1,forbidden2,forbidden3} from  stochastic, high-dimensional signals. Symbolic analysis has also shown to be useful for classifying different types of signals \cite{rosso,li_2008,rosso2,parlitz} and, in bivariate analysis, for inferring the direction of information flow \cite{ste_prl_2008,ste2_prl_2008}. The symbolic approach involves the transformation of a time-series, $x(t)$, into a sequence of symbols, $s(t)$, by using an appropriated codification rule. A significant advantage of symbolization is its reduced sensitivity to observational noise. Complexity measures have been proposed to characterize the resulting symbolic sequence, a very popular one being the {\em permutation entropy} \cite{bandt_PRL_2002,bkp_2002,keller_2003,cao_2004,lehnertz_2007,epileptic_2012,review}, computed with an ordinal symbolization rule, by comparing neighboring values in the time-series.

Complex networks have also been successfully employed for time-series analysis \cite{nicolis_2005,zhang_small_prl_2006,pnas_2008,physicad_2008,plos_one_2011,small_2014}. Correlation graphs \cite{yang_physicaA_2008,emilio}, recurrence graphs \cite{donner_NJP_2010,ewi1,ewi2} and visibility graphs \cite{lacasa_2008,lacasa_pre_2010,luque_plos_one_2011,luque_chaos_2012,hv_pre_2013} have been shown to provide relevant information, for example, of early-warning indicators of qualitative changes and abrupt transitions.

{\textcolor{black}{Here we characterise the evolution of a dynamical system by combining}} symbolic ordinal analysis and network representation, in a hybrid methodology. Specifically, a time series $x(t)$ (raw data) is transformed in a sequence, $s(t)$, of symbols (ordinal patterns, OPs) that is used to construct a {\em directed and weighted graph} where the different symbols that appear in $s(t)$ constitute the network nodes. Each of these nodes has an associated probability of occurrence. The transitions between consecutive symbols constitute the network links: any pair of nodes, $i$ and $j$, are connected by a directed link if the sequence of symbols $(i,j)$ occurs in the symbolic sequence, and the weight of the directed link is the probability of $j$ occurring after $i$ (referred to as the transition probability $i \rightarrow j$, TP).

{\textcolor{black}{Our approach for defining the links as the transition probability between symbols was originally proposed by Nicolis \emph{et al.} \cite{nicolis_2005} and the use of ordinal analysis was proposed by Sun \emph{et al.} \cite{small_2014}; however, in \cite{small_2014} the weights of the links were not defined in terms of TPs (as in \cite{nicolis_2005}), but rather, they were}} defined as the number of times the pair of consecutive symbols $(i,j)$ occurred in the time series. A major advantage of {\textcolor{black}{the use of TPs is that they}} are normalized in each node, and thus, allow for computing an entropy for each node, in the following referred to as \emph{node entropy}. The node entropies can then be used to identify changes in the network, which reflect changes in the time-series when the control parameter is varied.

Exploiting this network representation we propose several novel measures to characterize a time series:
\begin{itemize}
\item the {\em nodes' entropy}, which is the mean entropy of the distribution of the weights of the outgoing links of a node (in other words, is the node entropy, averaged over all the nodes of the network.);
\item the {\em links' entropy}, which is the entropy of the distribution of links' weights;
\item  the {\em asymmetry coefficient}, which is computed from the difference of the weights of the links $i \rightarrow j$ and $j \rightarrow i$, averaged over all the links of the network
\end{itemize}

\noindent Note that the nodes' and the links' entropies are computed according to the classical Shannon entropy definition.

By analyzing numerical data generated from the logistic map, the tent map, and the circle map, and also experimental data, recorded from the intensity of semiconductor lasers under different operation conditions, we show that these measures are able to capture gradual and abrupt changes in time-series. A comparison with the permutation entropy (PE) reveals that the entropies defined from the symbolic network vary over a wider range of values and outperform the permutation entropy in providing {\em early warning signals} of sudden changes in the time-series.

This paper is organized as follows. In Sections 2 and 3 the method of network construction and the network-based measures proposed for characterizing time-series are described. In Section 4 results are presented and, finally, we provide the discussion and some conclusions in Section 5.

\section{Network construction}

We transform a time series $x(t)$ into a sequence of symbols $s(t)$ by using the ordinal pattern (OP) representation with symbols of length $D$. In this case, symbols are defined by considering groups of $D$ consecutive values in the time series \cite{bandt_PRL_2002}. There exist $D$! different OPs of length $D$. For example, for $D=2$ there are two OPs: $x(t)<x(t+1)$ gives symbol `01' and $x(t)>x(t+1)$ gives symbol `10'; for $D=3$ there are six possible OPs: $x(t)<x(t+1)<x(t+2)$ gives `012', $x(t+2)<x(t+1)<x(t)$ gives `210', etc.

In general, the ordinal representation of $x(t)$ gives a sequence $s(t)$ of OPs with $M$ different symbols because depending on the system's dynamics not all possible symbols will be present in the symbolic sequence, $s(t)$, either because the dynamics does not allow for some of the (the so-called``forbidden patterns''), or because the time series $x(t)$ is of finite length and simply they do not appear (``missing patterns'')~\cite{forbidden1,forbidden2,forbidden3}. Thus, the number of nodes in the network corresponds only to those symbols appearing in the symbolic sequence and is $M \le D$!.

The weight of a node $i$ is the relative number of times the symbol $i$ occurs in the sequence $s(t)$:
\begin{equation}
  p_{i}=\frac{1}{L}\sum_{t=1}^L n[s(t)=i],
\label{pi}
\end{equation}
where $n$ is a count of the number of occurrences and $L$ is the length of the sequence $s(t)$. The weights are normalized, $\sum_{i=1}^M p_i =1$, and the permutation entropy, $s_{p}$ (PE), is the entropy of the distribution of node weights \cite{bandt_PRL_2002},
\begin{equation}
  s_{p} = -\sum_{i=1}^M p_i \log p_i.
\end{equation}
We note that $0\leq s_p \leq \log M$, with $s_p = 0$ if $p_i^*=1$ for $i=i^*$ and $p_i=0$ $\forall i \ne i^*$ (only one symbol, $i^*$, appears in the symbolic sequence and thus the network has only one node); $s_p = \log M$ if $p_i = 1/M$ $\forall i$ (all the symbols appear in the symbolic sequence with the same probability and thus all the nodes have the same weight).

The weight of a link $i\rightarrow j$, $w_{ij}$, is the relative number of times, in the sequence $s(t)$, the symbol $i$ is followed by the symbol $j$:
\begin{equation}
  w_{ij}=\frac{\ds\sum_{t=1}^{L-1} n[s(t)=i,s(t+1)=j]}{\ds\sum_{t=1}^{L-1} n[s(t)=i]}.
  \label{tp}
\end{equation}
With this definition the link weights are normalized in each node, i.e., $\sum_{j=1}^M w_{ij}=1,\ \forall i$. We note that this definition allows for the presence of self-loops.

\begin{figure}
\center\includegraphics[width=0.8\columnwidth,clip=]{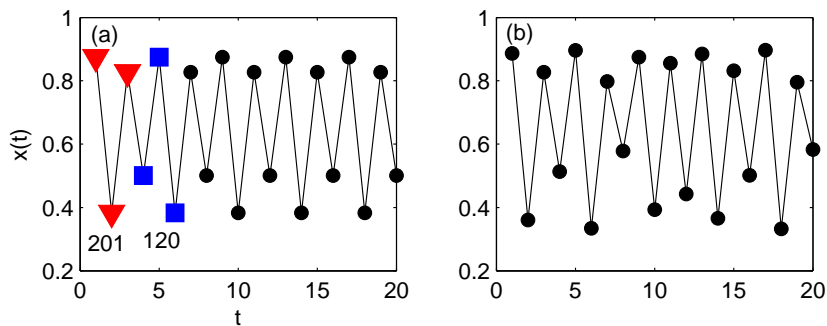}
\caption{Time series generated from the logistic map with (a) $r= 3.5$ and (b) $r=3.58$. In panel (a) the triangles and the squares indicate two ordinal patterns formed by two sets of $D=3$ consecutive and non-overlapping values, which give the transition $201\rightarrow 120$.}
\label{fig:nueva}
\end{figure}

It is illustrative to describe the type of networks that are generated by simple dynamics. A periodic time-series will give a regular symbolic sequence, whose periodicity will depend on the length of the ordinal pattern, $D$. The resulting symbolic network will then depend on $D$. To focus the ideas, let's consider the time series shown in Fig.~\ref{fig:nueva}(a), with period 4, generated from the logistic map. With $D=3$ we obtain the following sequence of symbols

\begin{equation}
\begin{array}{cccccccc}
201, & 021, & 102, & 120, & 201, & 021, & 102, &\dots .
\nonumber\end{array}
\end{equation}

Because the TPs are computed from ordinal patterns that are formed by non-superposed values [i.e., for $D=3$, \{$x(t)$, $x(t+1)$, $x(t+2)$\} define one pattern and \{$x(t+3)$, $x(t+4)$, $x(t+5)$\} define the next one], the network obtained has 4 nodes connected as follows:

\begin{equation}
\begin{array}{cccc}
201 \rightarrow 120, & 021 \rightarrow 201, & 102 \rightarrow 021, & 120 \rightarrow 102.
\end{array}\nonumber
\end{equation}

If we use $D=4$ OPs, then the sequence of symbols is
\begin{equation}
\begin{array}{cccccccc}
3021, & 0213, & 2130, & 1302, & 3021, & 0213, & 2130, &\dots
\end{array}\nonumber
\end{equation}

and the resulting network has 4 nodes with self-loops,
\begin{equation}
\begin{array}{cccc}
3021 \rightarrow 3021, & 0213 \rightarrow 0213, & 2130 \rightarrow 2130, & 1302 \rightarrow 1302.
\end{array}\nonumber
\end{equation}
It is important to note that, in the ordinal encoding scheme, depending on the value of $D$, an irregular time series can have associated a regular symbolic sequence, and thus, a regular network. To fix the ideas, let's consider the time series shown in Fig. \ref{fig:nueva}(b), also generated from the logistic map. With values of $D$ equal to 2, 3 or 4, this time-series gives the same symbolic sequence as the period-4 time-series shown in panel (a). This feature is a consequence of the ordinal codification rule. Other codification choices can result in two different symbolic sequences.

On the other hand, if the time-series is fully random, the symbolic sequence will also be fully random, and then the network will have a regular, all-to-all topology, regardless of the value of $D$ (assuming the time-series is sufficiently long).

%%%%%%%%%%%%%%%%%%%%%%%%%%%%%%%%%%%%%%%%%%%%%%%%%%%%%%%%%%%%
%%%%%%%%%%%%%%%%%%%%%%%%%%%%%%%%%%%%%%%%%%%%%%%%%%%%%%%%%%%%

\section{Network-based diagnostics for time-series analysis}

The symbolic network allows defining several novel measures:

1) Since $\sum_{j=1}^M w_{ij}=1,\ \forall i$, we can compute an entropy at each node,
\begin{equation}
  \label{node_entropy}
  s_i=-\sum_{j=1}^{M} w_{ij} \log w_{ij},
\end{equation}
and then characterize the network heterogeneity in terms of the distribution of the $s_i$ values.

We note that, on one hand, $s_i=0$ if and only if node $i$ has only one outgoing link, $j^*$: in this case, $w_{ij^*}=1$ and $w_{ij}=0$, $\forall$ $j\ne j^*$, and the symbol $j^*$ occurring after symbol $i$ is always the same. In this case, in the symbolic sequence, the symbol that follows $i$ is fully predictable. On the other hand, if all the outgoing links of node $i$ have the same weight, then $w_{ij}=1/k_i$ and $s_i=\log k_i$ (here $k_i$ is the out-degree of node $i$, $k_i=\sum_{j=1}^M A_{ij}$, where $A_{ij}$ are the components of the adjacency matrix, $A_{ij}=1$ if $w_{ij}\ne 0$, otherwise $A_{ij}=0$). The maximum $s_i$ value is $\log M$ and corresponds to a node that is connected to all the other nodes --including itself-- with links that have uniform weights. In this case, in the symbolic sequence any symbol can follow $i$ with the same probability.

We propose the use of the first moment of the distribution of $s_i$ values, the network-averaged node entropy,
\begin{equation}
  s_{n} = \frac{1}{M} \sum_{i=1}^M s_i,
\end{equation}
as a novel measure for the analysis of a time-series. We note that $0\leq s_n \leq \log M$. If $s_n=0$, then $s_i=0$ $\forall i$, and all the nodes have only one outgoing link; therefore, the symbol $j$ that occurs after symbol $i$ is fully predictable, $\forall i$. On the contrary, the largest $s_n$ value, $\log M$, occurs when $s_i=\log M$ $\forall i$. In this case the symbolic sequence is fully random and after any symbol, $i$, any symbol in the sequence can follow with the same probability $1/M$.

We note that the range of variation of the node entropies, $s_i$, and of their average, $s_n$, is the same as that of the permutation entropy, $s_p$.

2) From the probability distribution function (pdf) of the weights of all the links, $p(w)$, we define the {\em link's entropy} as:

\begin{equation}
  s_{l} =-\int_0^1 p(w)\log p(w) dw.
\end{equation}

We have estimated $p(w)$ from an histogram of the weights of the existing links ($w_{ij}\ne 0$). Large values of $s_l$ indicate a heterogeneous distribution of link weights, while small values of $s_l$ indicate a narrow distribution of link weights. The delta distribution, $w_{ij}=w$, $\forall i,j$ that gives $s_l=0$ includes two limits, one in which the symbolic sequence is fully predictable (each node has only one outgoing link: $\forall i$, $w_{ij_i^*}=1$ and $w_{ij}=0$ if $j\ne j_i^*$) and the other in which the symbolic sequence is fully unpredictable (the nodes are all-to-all connected with uniform weights: $w_{ij}=1/M$, $\forall i,j$). Thus, $s_l$ quantifies the \emph{complexity} of the symbolic sequence, as $s_l=0$ corresponds to both, perfectly regular and fully random sequences.

3) To quantify the asymmetry in the direction of the network links, we introduce the {\em asymmetry coefficient}, which is defined as:

\begin{equation}
  a_c = \frac{\ds\sum_i \sum_{j\ne i} |w_{ij}-w_{ji}|}{\ds\sum_i \sum_{j\ne i} (w_{ij}+w_{ji})}\,.
\end{equation}
$a_c=0$ in a fully symmetric network ($w_{ij}=w_{ji}$, $\forall i,j$) and $a_c=1$ in a fully directed network (either $w_{ij}=0$ or $w_{ji}=0$ $\forall i,j$). This asymmetry coefficient is closely related to the concept of reciprocity in graphs \cite{garla2004}.

We note that for computing these measures, only the nodes and the links that are present in the network are taken into account. While absent nodes and links, corresponding to missing or forbidden symbols and transitions, could have been taken into account with zero weights, we preferred to use this more generic approach because, depending on the symbolic transformation used, the complete set of symbols can be unknown or even infinite.

\section{Results}

\subsection{Analysis of synthetic data}

We start by presenting the results of the analysis of simulated time-series for the logistic map, the tent map and the stochastic circle map. The equations and parameters are

\begin{itemize}
\item logistic map: $x_{i+1}=rx_{i}(1-x_i)$;
\item tent map: $x_{i+1}=rx_{i}$ if $x_i<0.5$, $x_{i+1}=r(1-x_{i})$ if $x_i\geq 0.5$;
\item stochastic circle map: $\phi_{i+1} = \phi_i+ \rho + k (\sin 2\pi\phi_i)/(2\pi) +\beta \xi$. Here $\xi$ is a Gaussian white noise with unit standard deviation, $k$ is the map parameter, $\beta$ is the noise strength and $\rho=-0.23$ is kept fixed. We analyze time series of $x_i=\phi_i-\phi_{i-1}$.
\end{itemize}

The analysis is performed with time-series of length {\textcolor{black}{$L=6000$}} and patterns of length $D=4$. The results are robust as long as $L$ is much larger than the number of possible links. The value of $D$ is chosen because it allows constructing symbolic networks with 24 possible nodes and 576 possible links; in contrast, $D=5$ gives networks with 120 possible nodes and 14400 possible links, and therefore, computing the links' weights with robust statistics requires long time-series. We discuss below the influence of $D$ and $L$, when we analyze empirical data.

\begin{figure}[tbh]
  \begin{tabular}{ll}
  \includegraphics[width=0.48\columnwidth,clip=]{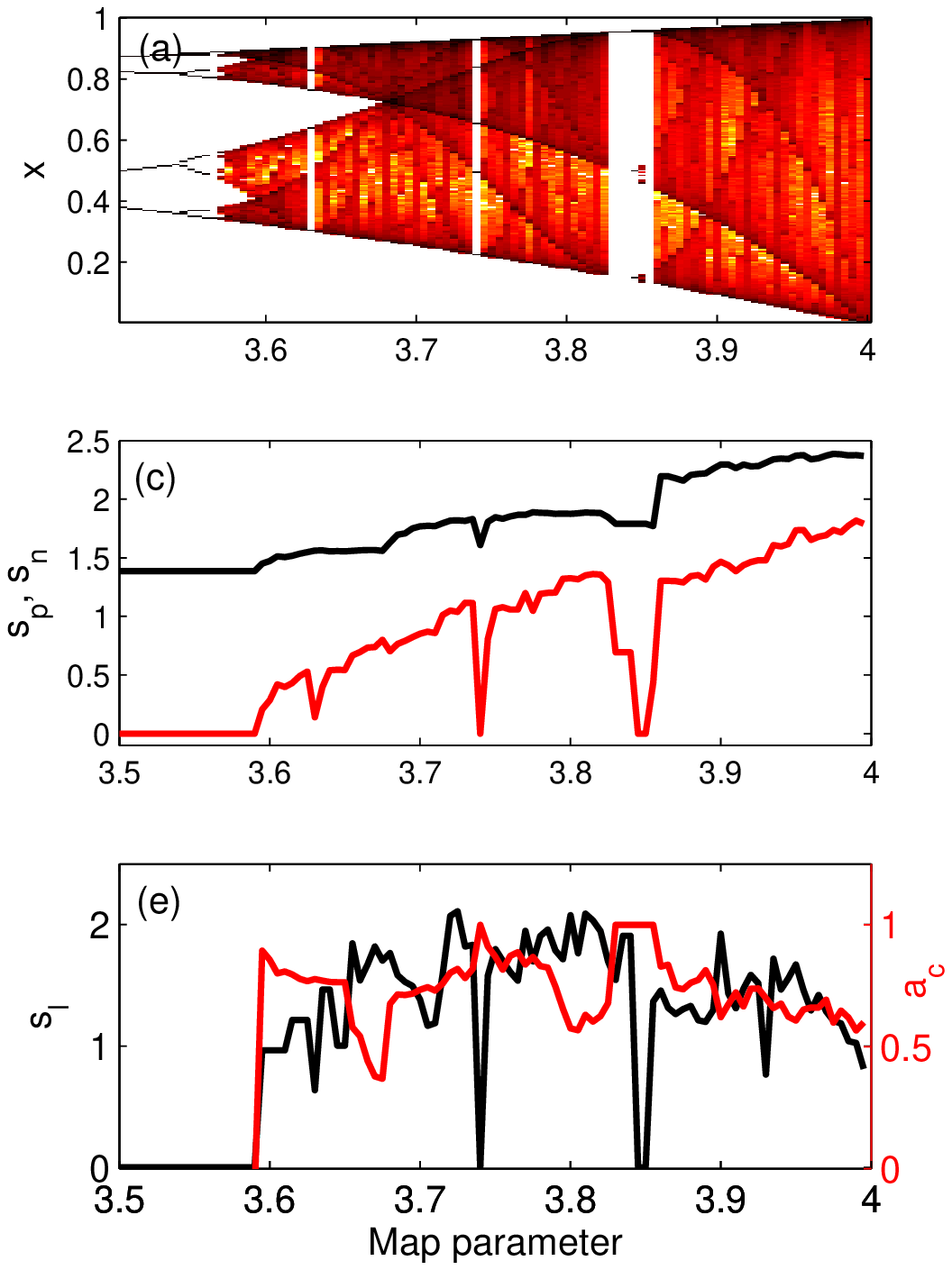} &
  \includegraphics[width=0.48\columnwidth,clip=]{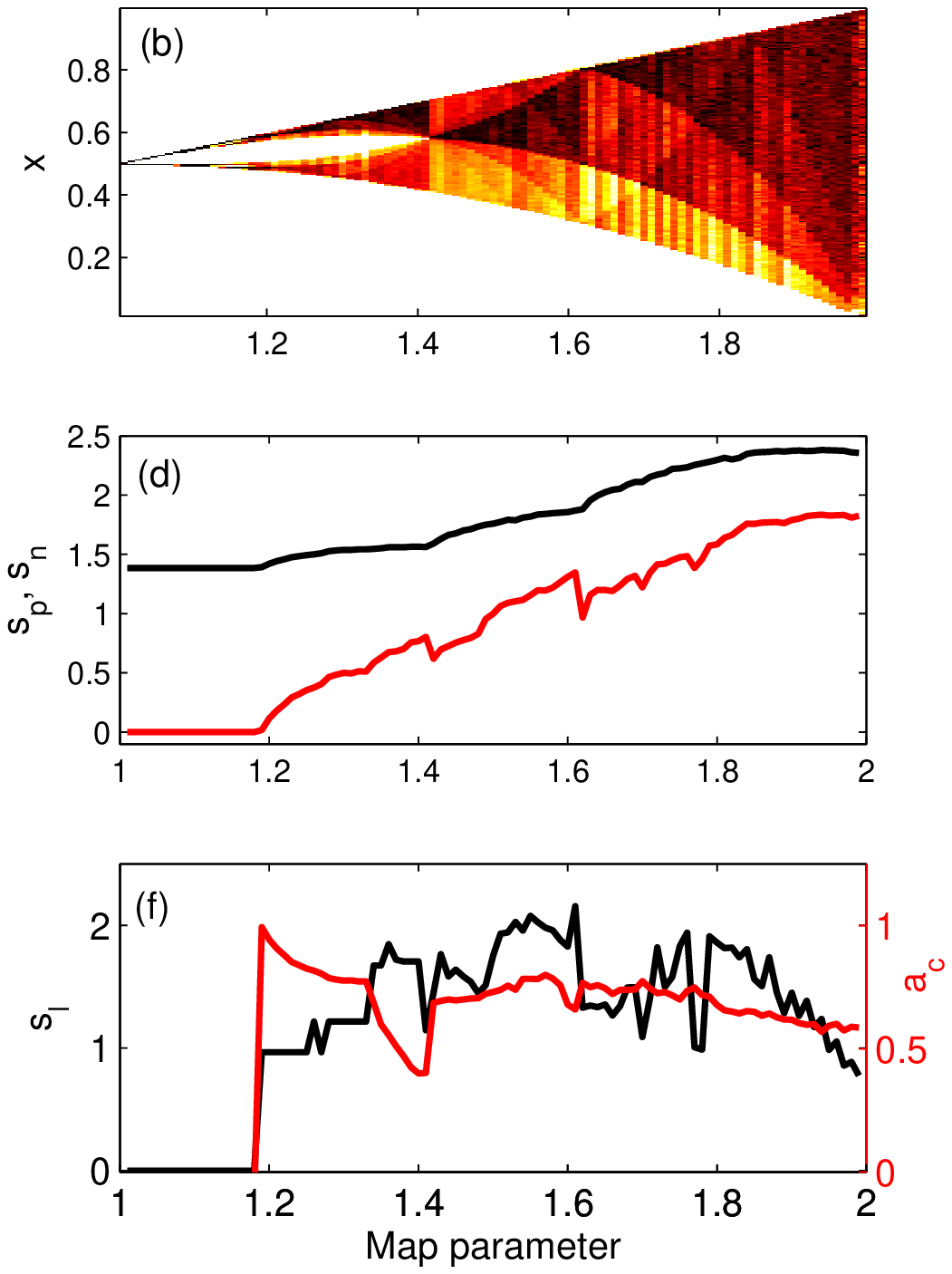}
  \end{tabular}
  \caption{Analysis of the logistic map (left column) and tent map (right column). (a), (b) Probability distribution function (pdf) of the variable $x$ (in color scale) vs. the map parameter. (c), (d) the average node entropy, $s_n$ (red line), and the permutation entropy, $s_p$ (black line). Note that the range of variation of $s_n$ is more than twice that of $s_p$. (e), (f) the links' entropy, $s_n$ (black line), and the asymmetry coefficient, $a_c$ (red line). The analysis was performed with $D=4$ and $L=6000$.}
  \label{fig:todos0}
\end{figure}

Regarding the logistic and tent maps, they have similar bifurcation sequences (except for the periodic windows in the logistic map) and thus, assuming similar bifurcation sequences give rise to similar dynamical behaviors, we expect to obtain similar symbolic networks, which will depend on the control parameter. Indeed, this is observed in Fig.~\ref{fig:todos0} that displays, for the logistic map (left column) and for the tent map (right column), the bifurcation diagrams (top row), the permutation entropy, $s_p$, and the average node entropy, $s_n$, (middle row) and the links' entropy, $s_l$, and the asymmetry coefficient, $a_c$ (bottom row). For both maps $s_p$ and $s_n$ increase in a similar way with the map parameter (except in periodic windows). One can note that $s_n$ varies over a wider range of values in comparison to $s_p$. We note that this difference in the actual range of variation of $s_n$ and $s_p$ can not simply be normalized over because, as discussed in Sec. 3, both, $s_n$ and $s_p$ vary within the same range of values. Also, $s_n$ displays more abrupt variations, which can be understood in the following terms: as the map parameter increases, the dynamics becomes increasingly chaotic and new OPs appear in the symbolic sequence, which result in new nodes and links in the symbolic network; however, the frequency of occurrence of the new OPs is initially small, and their appearance does not produce abrupt variations in the permutation entropy. But the new nodes will not necessarily have small entropies; therefore, they can induce abrupt variations in the average node entropy.

Both, $s_p$ and $s_n$ vary with the map parameter in a qualitatively similar way as the Lyapunov exponent (LE). For the logistic map it has previously been shown that for chaotic parameters and low $D$ values, $s_p$ behaves as the LE \cite{bandt_PRL_2002}. In addition, for piecewise monotone interval maps, in the limit of $D\rightarrow \infty$, $s_p$ tends to the topological entropy \cite{bkp_2002}.
In the visibility graph, a direct relationship between the network entropy and the LE can be established by using the Pesin identity \cite{luque_chaos_2012}. Therefore, these results suggest that the network entropy can indeed be expected to capture nontrivial properties of the dynamics; however, determining a direct correspondence between different network properties and different kinds of dynamics is not straightforward and is left for future work.

\begin{figure}[tbh]
  \begin{tabular}{ll}
  \includegraphics[width=0.48\columnwidth,clip=]{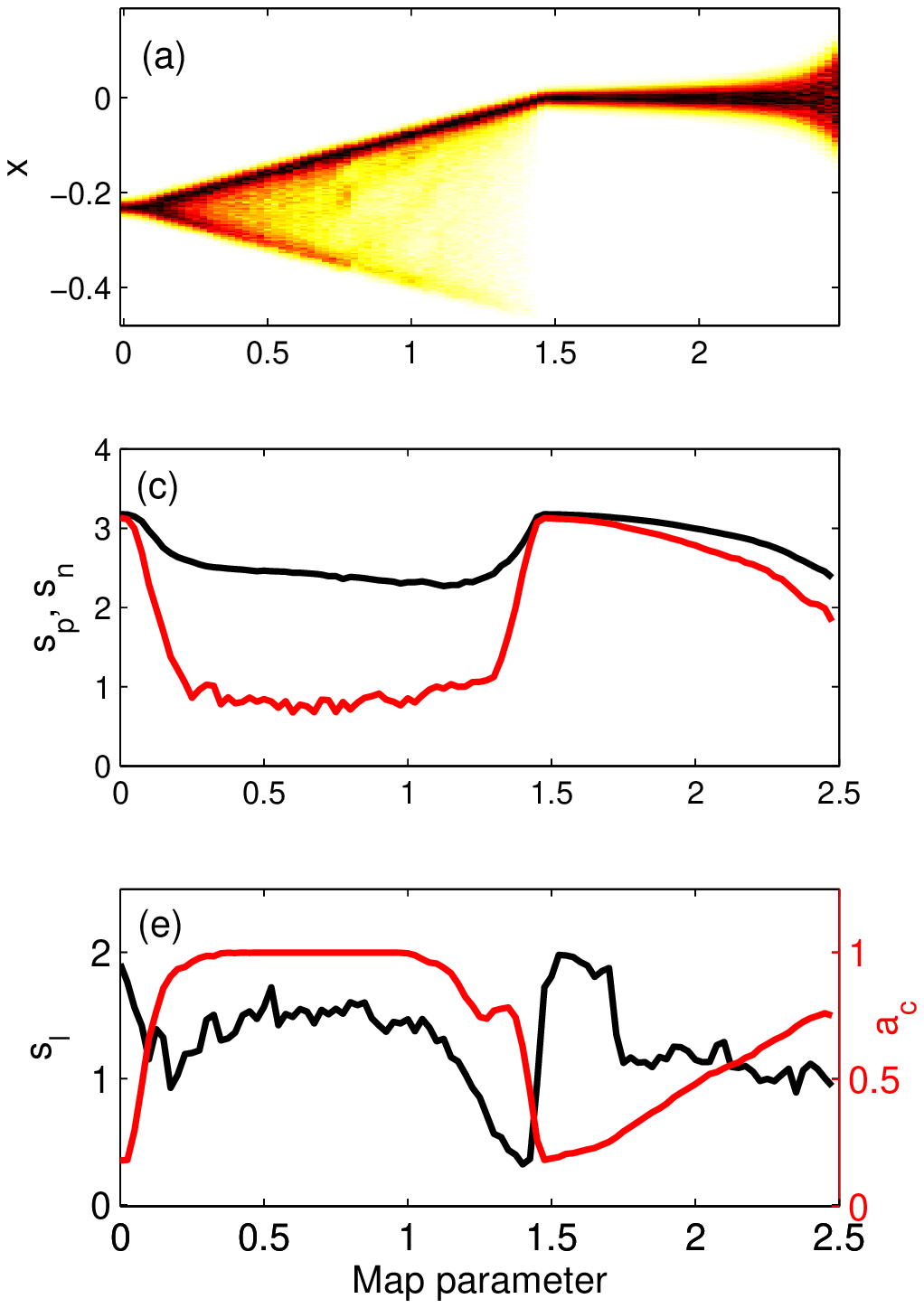} &
  \includegraphics[width=0.48\columnwidth,clip=]{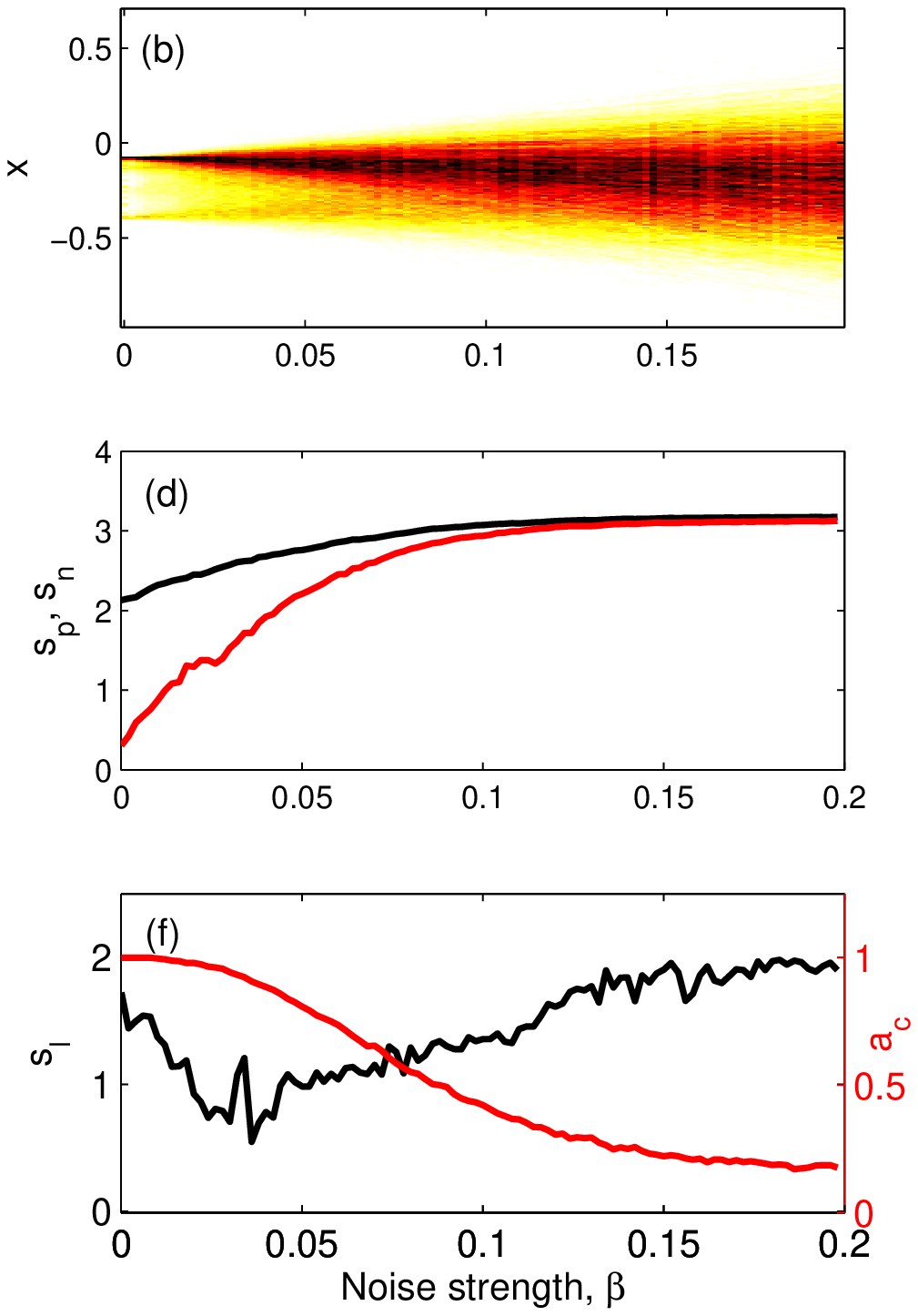}
  \end{tabular}
  \caption{{\textcolor{black}{As Fig.~\ref{fig:todos0} but for the stochastic circle map when the map parameter, $k$, is varied (left column) and when the noise strength, $\beta$, is varied (right column). }} }
  \label{fig:todos1}
\end{figure}

Figure~\ref{fig:todos1} displays the results of the analysis of simulations of the stochastic circle map, when varying the parameter that represents the forcing amplitude (left column) or the noise strength (right column). Varying $k$ allows to study the transition to locking. For very small values of $k$ the stochastic term dominates and the network contains all possible nodes and links, thus, $s_p$ and $s_n$ are maximum ($\simeq \log 4! = 3.18$). As $k$ increases the dynamics becomes more deterministic, the number of nodes and links gradually decrease (not shown) and $s_p$ and $s_n$ also decrease. At $k\simeq 1.5$ the transition to locking occurs [see the bifurcation diagram in Fig.~\ref{fig:todos1}(a)] and the presence of noise results in a symbolic sequence that is again fully random, thus, $s_p$ and $s_n$ again reach the maximum value. If $k$ is further increased, again an increase in the ``order'' occurs and $s_p$ and $s_n$ again decrease. The right column in Fig.~\ref{fig:todos1} reveals that $s_p$ and $s_n$ capture the monotonic increase of ``disorder'' in the symbolic sequence as the noise strength, $\beta$, increases.

%%%%%%%%%%%%%%%%%%%%%%%%%%%%%%%%%%%%%%%%%%%%%%%%%%%%%%%%%%%%
%%%%%%%%%%%%%%%%%%%%%%%%%%%%%%%%%%%%%%%%%%%%%%%%%%%%%%%%%%%%

Other network measures are consistent with these observations. As seen in Figs.~\ref{fig:todos0}(e) and~\ref{fig:todos0}(f), for the logistic and for the tent maps, when the dynamics is weakly chaotic ($r\approx 3.6$ for the logistic map; $r\approx 1.2$ for the tent map), the symbolic network has a low link entropy and is highly asymmetric (the asymmetry coefficient is close to 1); as the parameter increases the network grows and the link's entropy increases, accompanied by large variations of the asymmetry coefficient; for large parameter values (strong chaos) both, the asymmetry coefficient and the links' entropy decrease, suggesting that the network becomes more homogeneous. These results are consistent with a previous analysis of the complexity of the logistic map, where the peak values of the various complexity measures analyzed occurred at intermediate values of the map parameter \cite{rosso_physica_a_2006}. For the stochastic circle map,  $s_l$ and $a_c$ [Figs.~\ref{fig:todos1}(e) and~\ref{fig:todos1}(f)] reveal that the network is homogeneous (large link entropy accompanied by small asymmetry coefficient) for $k\simeq0$, at the locking transition, and for strong noise.

%%%%%%%%%%%%%%%%%%%%%%%%%%%%%%%%%%%%%%%%%%%%%%%%%%%%%%%%%%%%
%%%%%%%%%%%%%%%%%%%%%%%%%%%%%%%%%%%%%%%%%%%%%%%%%%%%%%%%%%%%

\subsection{Analysis of empirical data}

\begin{figure}[tbh]
  \includegraphics[width=1.0\columnwidth,clip=]{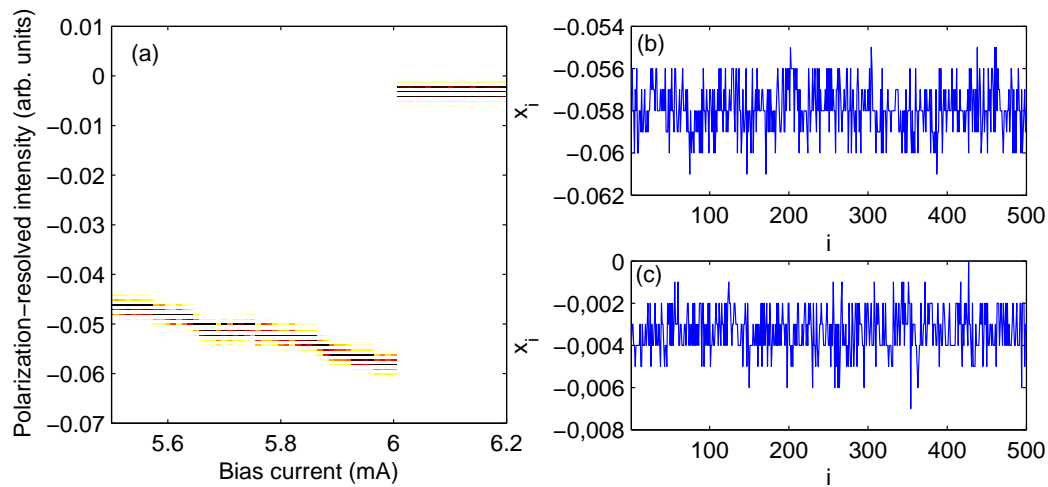}
  \caption{(a) Pdf of the intensity of the linear polarization mode that turns on vs. the laser bias current. Intensity time-traces just before (b) and after (c) the switch.}
  \label{fig:yanhua}
\end{figure}

Let us next present the results of the analysis of two empirical data sets. The first data set was recorded from the output of a semiconductor laser at various values of the laser bias current ranging from 5.5~mA to 6.2~mA. The type of laser is a vertical-cavity surface-emitting laser (VCSEL) that can emit two orthogonal polarizations and, when varying the bias current, there is an abrupt polarization switch (PS) at about 6~mA: the dominant polarization turns off and the orthogonal one turns on, as shown in Fig.~\ref{fig:yanhua} (a). The data consist of 71 time-series of the intensity of one polarization mode recorded at fixed values of the bias current, below and above the polarization-switching point. Each time series contains more than $10^6$ data points. In order to identify clear trends, each time-series was divided in several sections, each containing $L$ data points ($L$ varying in the range $10^3-10^5$ as discussed below), and the network indicators were computed by averaging the values in each section.

Since the empirical data is very noisy, {\textcolor{black}{see Figs.~\ref{fig:yanhua}(b) and~(c)}}, all possible pairs of symbols ($i,j$) occur in the symbolic sequence and the network is a regular, all-to-all graph. Nevertheless, we will show that the network measures adequately capture dynamical changes in the time-series.

%%%%%%%%%%%%%%%%%%%%%%%%%%%%%%%%%%%%%%%%%%%%%%%%%%%%%%%%%%%%
%%%%%%%%%%%%%%%%%%%%%%%%%%%%%%%%%%%%%%%%%%%%%%%%%%%%%%%%%%%%

\begin{figure}[tbh]
  \includegraphics[width=0.98\columnwidth,clip=]{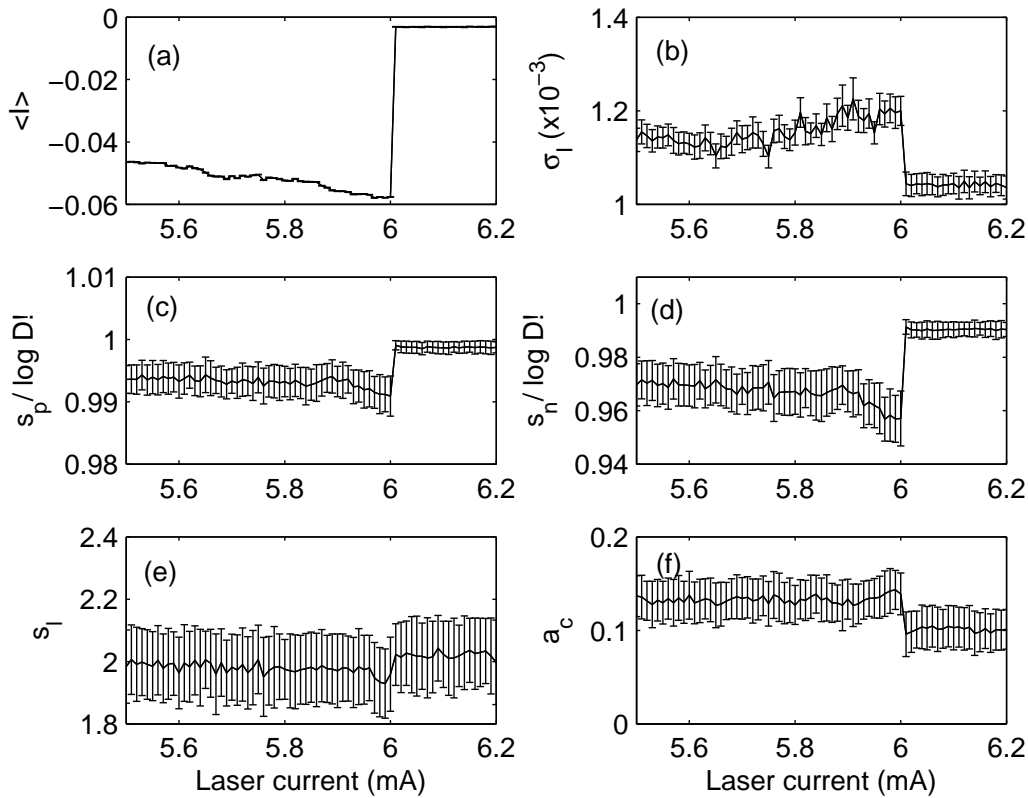}
  \caption{Early indicators of abrupt polarization switching (PS). The data analyzed is as in Fig.~\ref{fig:yanhua}. (a) the mean value and (b) the standard deviation of the intensity fluctuations; (c) the permutation entropy, (d) the average node entropy, (e) the links entropy and (f) the asymmetry coefficient. $s_p$ and $s_n$ are normalized to the maximum value, $\log D!$. The analysis was performed with $D=3$. The error bars indicates the standard deviation of the distribution of values computed over 100 sections with $L=1000$ data points each (see text for details). It can be seen that $s_p$, $s_n$ and $a_c$ are good indicators of the abrupt switching, as they change the linear slope before the PS; in contrast $\langle I \rangle$, $\sigma_I$ and $s_l$ are not good indicators as they either vary linearly or fluctuate with constant amplitude. $s_p$ varies nonlinearly before the PS but in a smaller range of values as compared with $s_n$. }
  \label{fig:yanha_estadistica}
\end{figure}

Figure~\ref{fig:yanha_estadistica} presents a comparison of the the network-based measures, with the mean value, $\langle I\rangle$, and the standard deviation, $\sigma$, of the intensity fluctuations: $\langle I\rangle$ and $\sigma$ vary linearly, and thus, they don't provide an early warning for the sudden polarization switch. It can also be seen that both, $s_p$ and $s_n$, change the slope of initial linear trend before the PS, but the $s_n$ trend decreases faster, providing a better indicator of the PS.

While the evidence of early warning in terms of a single network measure is not strong, taken together these results provide clear early warnings of abrupt switching. This suggests that the switching involves deterministic light-matter interactions that, despite of the highly stochastic character of the signals analyzed, the network measures are able to capture. In VCSELs the two linear polarizations are strongly anti-correlated and the polarization switching can be in part due to thermal effects (Joule heating) through a shift of the gain maximum \cite{book}. Also, small cavity anisotropies can play a role, as the VCSEL circular transverse geometry provides no polarization selection mechanism. In addition, material birefringence, saturable dispersion and a microscopic spin-flip relaxation mechanism can explain the PS \cite{dataexp}. Because various sources of noise are ubiquitous in laser systems (spontaneous emission noise, thermal and electrical noise), it is not possible to investigate experimentally the relative impact of stochastic and deterministic mechanisms that trigger polarization switchings. Various models suggest that several laser parameters can modify the linear stability of the two polarization modes, while small variations (deterministic or stochastic) are responsible for triggering the switch.

\begin{figure}[tbh]
  \includegraphics[width=0.98\columnwidth,clip=]{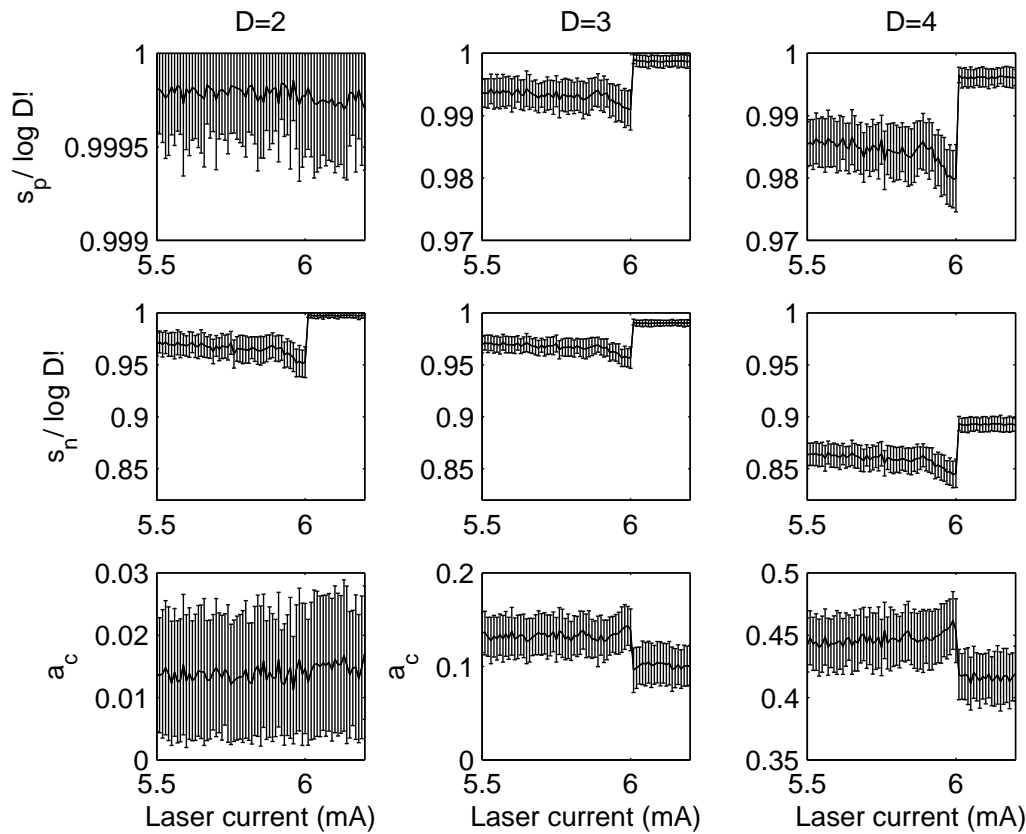}
  \caption{Influence of the length $D$ of the ordinal pattern. The data analyzed is as in Fig.~\ref{fig:yanhua}. The left column presents results for $D=2$, the central column, for $D=3$ and the right column, for $D=4$. The top row displays the permutation entropy, the central row, the average node entropy, and the bottom row, the asymmetry coefficient. For easy comparison $s_p$ and $s_n$ are normalized to the maximum value, $\log D!$. The error bars indicates the standard deviation of the distribution of values computed over 100 sections with $L=1000$ data points each (see text for details). The behavior of the network measures is robust for $D=3$ and $D=4$ and even for $D=2$, $s_n$ anticipates the polarization switching.}
  \label{fig:yanhua_d}
\end{figure}

\begin{figure}[tbh]
  \includegraphics[width=0.98\columnwidth,clip=]{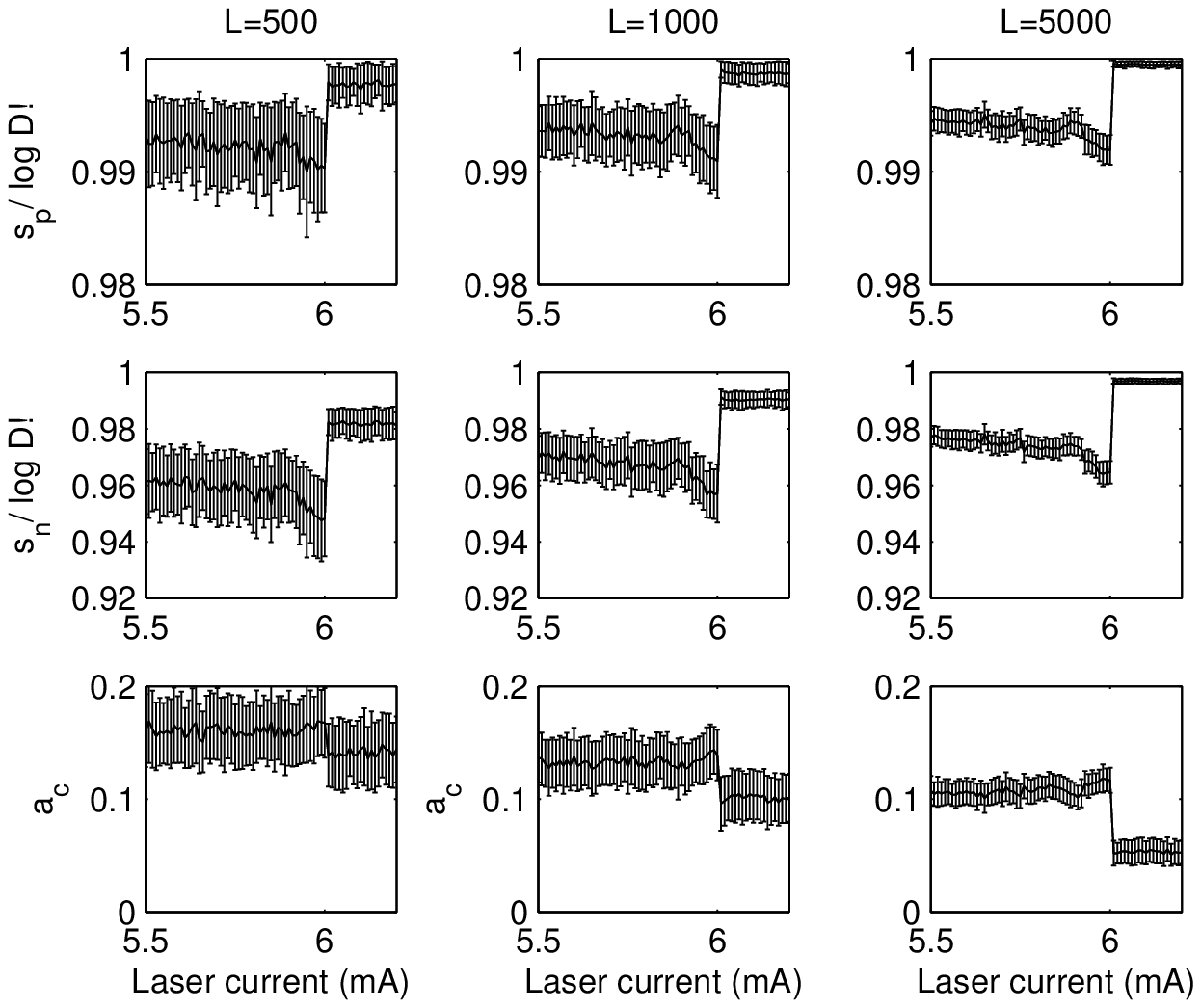}
  \caption{As in Fig.~\ref{fig:yanhua_d} but now we analyze the effect of the length of the time-series: in the left column, $L=500$; in the central column, $L=1000$ and in the right column, $L=5000$. In each panel $D=3$. We note that, as the length of the data set increases, the size of the error bars diminishes.}
  \label{fig:yanhua_l}
\end{figure}

Figures~\ref{fig:yanhua_d} and~\ref{fig:yanhua_l} analyze the influence of the length of the ordinal pattern, $D$, and the number of data points, $L$. In Fig.~\ref{fig:yanhua_d} we note that the above observations are robust with respect to $D$, while in Fig.~\ref{fig:yanhua_l}, we note that if $L$ is too short, the network measures fail to provide an early warning.

%%%%%%%%%%%%%%%%%%%%%%%%%%%%%%%%%%%%%%%%%%%%%%%%%%%%%%%%%%%%
%%%%%%%%%%%%%%%%%%%%%%%%%%%%%%%%%%%%%%%%%%%%%%%%%%%%%%%%%%%%

%%%%%%%%%%%%%%%%%%%%%%%%%%%%%%%%%%%%%%%%%%%%%%%%%%%%%%%%%%%%
%%%%%%%%%%%%%%%%%%%%%%%%%%%%%%%%%%%%%%%%%%%%%%%%%%%%%%%%%%%%

%%%%%%%%%%%%%%%%%%%%%%%%%%%%%%%%%%%%%%%%%%%%%%%%%%%%%%%%%%%%
%%%%%%%%%%%%%%%%%%%%%%%%%%%%%%%%%%%%%%%%%%%%%%%%%%%%%%%%%%%%

\begin{figure}[tbn]
  \includegraphics[width=0.8\columnwidth,clip=]{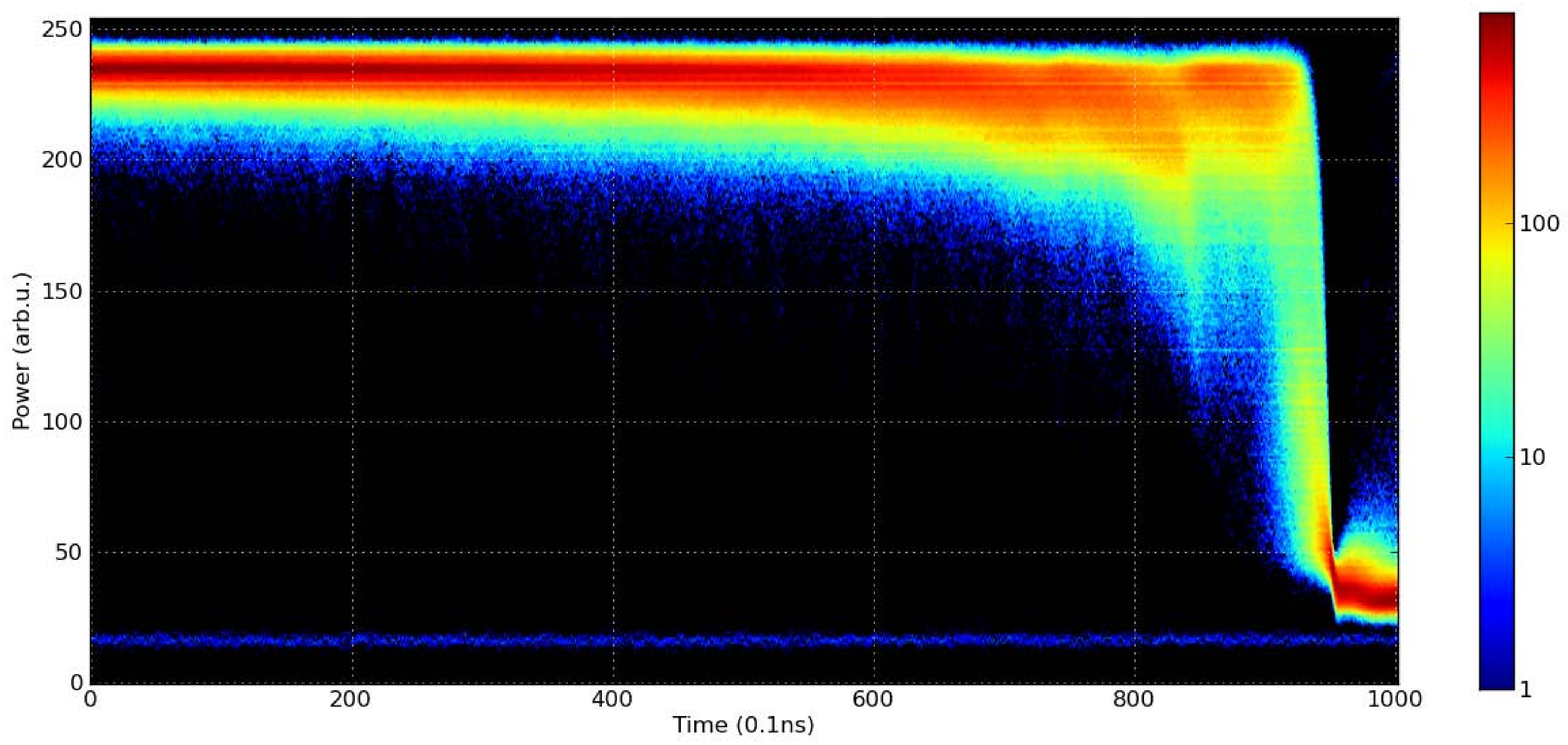}
  \caption{Second empirical data set recorded from a semiconductor laser with orthogonal optical feedback: pdf of the intensity (in color code) of the linear polarization mode that turns off vs. time. The pdf is computed by averaging over 1000 time-series.}
  \label{fig:new}
\end{figure}

To confirm these observations we now consider a second empirical data set, in which the intensity of one polarization mode was recorded while the control parameter, the laser pump current, was linearly scanned across the switching point. The data was recorded from the output of a different VCSEL, which was subject to polarized optical feedback. In this VCSEL the polarization switching (shown in Fig.~\ref{fig:new}) occurred only if the feedback was strong enough; without feedback no polarization switching was observed in this device. This setup has therefore the advantage that, by finely tuning the feedback strength, we could record two sets of time-series: one set in which a PS occurs, and, by decreasing slightly the feedback strength, another set in which no PS occurs.

We recorded 1000 independent realizations (each time-series has 20000 data points), and in each time-series the control parameter, the laser current, varied from a value below the PS to a value above the PS (see Fig.~\ref{fig:new}). For comparison, a second data set (also of 1000 time-series with 20000 data points) was recorded with slightly weaker feedback, such that no switchings were observed.

We divided each time-series in non-overlapping sections and computed the network measures in each section. As we are interested in anticipating the switching in real time, we want to consider sections as short as possible, but on the other hand, they can not be too short as we need to compute the weights of the links with good statistics. We found that using sections of $L=500$ data points each, and ordinal patterns of $D=3$, provided a good compromise, because we have 6 nodes, 36 links and their probabilities can be computed, in each section, with sufficient statistics.

\begin{figure}[tbn]
  \includegraphics[width=0.8\columnwidth,clip=]{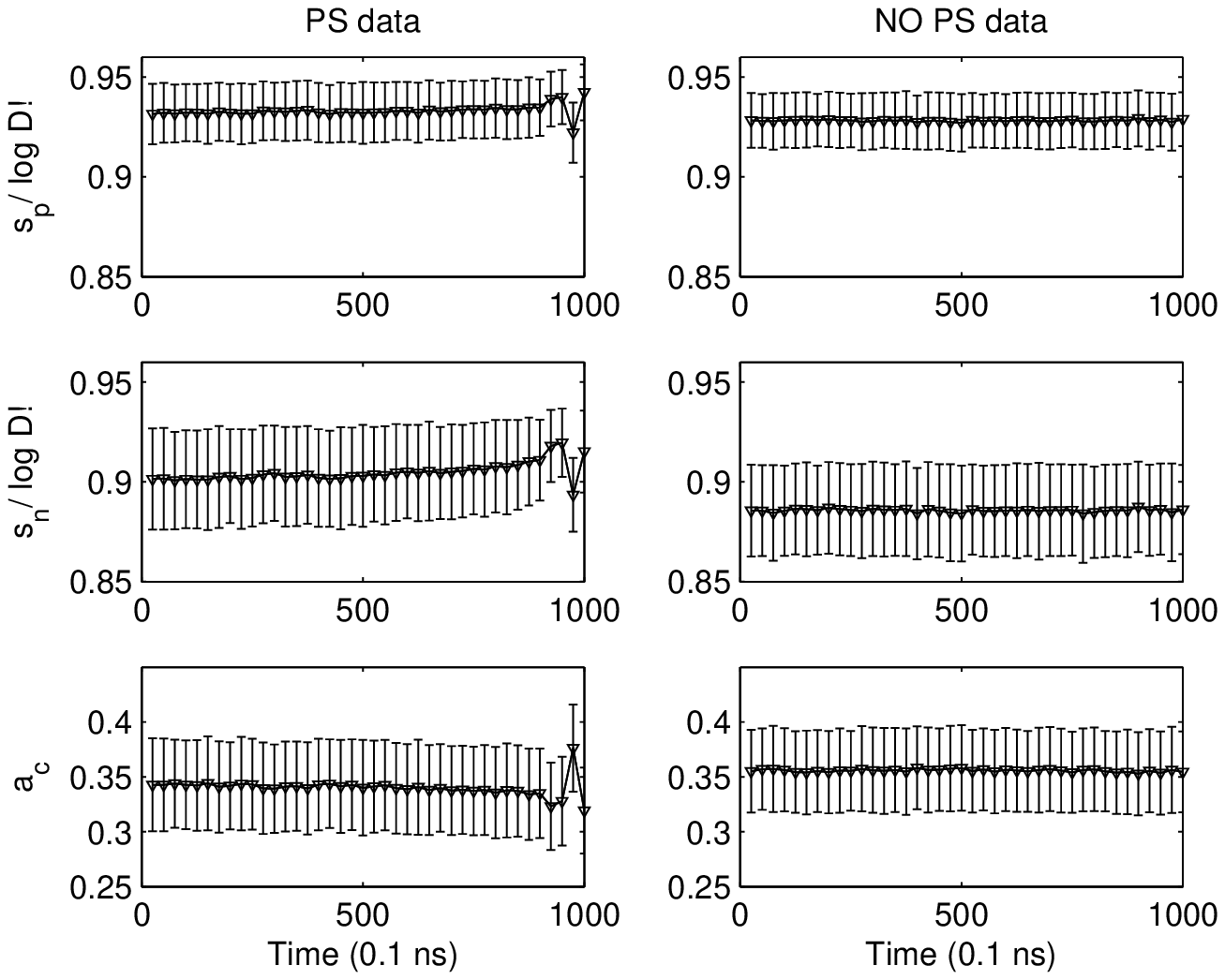}
  \caption{Comparison between 1000 independent realizations in which polarization switchings occur (left column) and 1000 independent realizations in which no switching occurs (right column, see text for details). Top row: the permutation entropy; middle row: the average node entropy; bottom row: the asymmetry coefficient, $a_c$. The analysis is performed with $D=3$, and each time series is divided in 40 non-overlapping sections of $L=500$ data points each. $s_p$ and $s_n$ are normalized to the maximum value, $\log D!$. While the dispersion of values is quite large (due to the short length of each section), in the left column there is a clear trend in the distributions of $s_p$ and $s_n$ values when approaching the switching point. In comparison the variation of $a_c$ is small and thus, $a_c$ does not constitute a good indicator of the PS. We observe that $s_n$ varies over a wider range of values with respect to $s_p$ and, when approaching the switching, both $s_n$ and $s_p$ increase reaching similar values.}
  \label{fig:nice2}
\end{figure}

The results are presented in Fig.~\ref{fig:nice2}, which displays de distribution of $s_p$, $s_n$ and $a_c$ values computed over the 1000 independent realizations: the left column corresponds to the data set in which PS occurs, while in the right column, no PS occurs. We observe that $s_n$ varies over a wider range of values with respect to $s_p$ and, when approaching the switching, both $s_n$ and $s_p$ increase reaching similar values. The asymmetry coefficient decreases only slightly when approaching the PS and thus is not a good indicator of the switching ahead (this is probably because of the short section length, see the influence of $L$ in the analysis of the first empirical data set, Fig. \ref{fig:yanhua_l}).

%%%%%%%%%%%%%%%%%%%%%%%%%%%%%%%%%%%%%%%%%%%%%%%%%%%%%%%%%%%%
%%%%%%%%%%%%%%%%%%%%%%%%%%%%%%%%%%%%%%%%%%%%%%%%%%%%%%%%%%%%

%%%%%%%%%%%%%%%%%%%%%%%%%%%%%%%%%%%%%%%%%%%%%%%%%%%%%%%%%%%%
%%%%%%%%%%%%%%%%%%%%%%%%%%%%%%%%%%%%%%%%%%%%%%%%%%%%%%%%%%%%

\section{Conclusions and discussion}

{\textcolor{black}{We have used}} a hybrid methodology to characterize the evolution of a dynamical system that combines two mathematically grounded tools for time-series analysis: symbolic ordinal analysis and network representation. Specifically, time series were transformed in sequences of ordinal patterns that were used to construct directed and weighted graphs. Exploiting this graph representation we propose several novel diagnostics to characterize time series: the averaged node entropy, the entropy of the distribution of links' weights, and the asymmetry coefficient. The analysis of numerical data (generated from the logistic, tent and circle maps) and experimental data (recorded from the output of semiconductor lasers under different external perturbations) demonstrated that these network-based measures are suitable for characterizing time-series, adequately capturing the effects of parameter changes that result in subtle or sudden transitions.

The links entropy is an interesting quantity for measuring the complexity of the symbolic sequence associated to a time-series, because it is equal to 0 when the weights of the links are delta-distributed, which occurs when the symbolic sequence is either perfectly regular or fully stochastic. In agreement with a previous study that found maximum complexity of the logistic map at intermediate values of the map parameter \cite{rosso_physica_a_2006}, here we found that the links' entropy is maximum at intermediate values of $r$.

The analysis of the empirical data confirmed that network-based measures provide early warning signals of abrupt transitions, despite of the highly stochastic character of the signals analyzed. These network-based networks can thus complement other indicators, such as the standard deviation or the permutation entropy. The methodology proposed here can be a valuable tool for the analysis of a wide range of systems where critical transitions might occur, including population extinctions, desertification, wetland degradation and epileptic seizures, to name a few.

Our approach could allow identifying changes in the symbolic dynamics of these systems, for example, the appearance of new symbols or the appearance of new transitions, which result in variations of the network measures. Bifurcations from a chaotic attractor to another chaotic attractor could be identified if the two attractors have associated symbolic dynamics which have different statistical distribution of symbols and transition probabilities. On the contrary, if a bifurcation does not result in changes in the symbolic dynamics of the system, the network measures proposed here won't be able to capture the bifurcation (see, e.g., the two time-series displayed in Fig.~\ref{fig:nueva}, which, with $D$ being 2, 3 or 4, ordinal patterns are encoded in the same symbolic sequence).

{\textcolor{black}{Another important drawback is that both, periodic and irregular time series could eventually be mapped to the same class of regular networks. Note that this methodology is not intended to classify between different time series, but to detect changes in the time series as a function of the tuning parameter. Thus, this methodology should be only used to complement other conventional tools of time-series analysis if the goal is a classification. In spite of these drawbacks}}, our approach provides additional predictive power for time-series analysis, as identifying the nodes which have one (or a reduced number) of outgoing links will allow identifying ``predictable symbols'' in the symbolic sequence, those for which we know which is the next most probable symbol in the symbolic sequence. If ordinal analysis is used to construct the symbolic network, then the information gained about the next most probable symbol will not allow inferring the variations of values in the time-series, but it will allow inferring the shape of the future oscillation in the time-series. In this sense, the symbolic network-based analysis proposed here will provide complementary information to that gained by using other well-established tools for time-series analysis.

This work has been supported by the Spanish MINECO (FIS2012-37655-C02-01 to CM, FIS2012-38266 to AA and SG), the Generalitat de Catalunya (ICREA Academia to CM and AA), EOARD (Grant FA9550-14-1-0359) to CM, the European Commission FET-Proactive Project MULTIPLEX (Grant 317532) to AA and SG, and the James S.\ McDonnell Foundation to AA.
%%%%%%%%%%%%%%%%%%%%%%%%%%%

%====================================================================
\end{document}